\documentclass[]{aastex}
\usepackage{emulateapj5}

\usepackage{psfig}

\slugcomment{Accepted for publication in ApJ, part 1}

\begin{document}

\title{The jet and circumburst stellar wind of GRB~980519
\altaffilmark{1}
}

\author{A.~O.~Jaunsen\altaffilmark{2,3},
 J.~Hjorth\altaffilmark{4,3,5},
 G.~Bj\"ornsson\altaffilmark{6},
 M.~I.~Andersen,\altaffilmark{7,8}
 H.~Pedersen\altaffilmark{4},
 K.~Kjernsmo\altaffilmark{2},
 H.~Korhonen\altaffilmark{7,8},
 P.~M.~S\o rensen\altaffilmark{9},
 E.~Palazzi\altaffilmark{10}}

\altaffiltext{1}{The Nordic Optical Telescope is operated on the
island of La Palma jointly by Denmark, Finland, Iceland, Norway, and
Sweden, in the Spanish Observatorio del Roque de los Muchachos of the
Instituto de Astrofisica de Canarias. The data presented here have
been taken using ALFOSC, which is owned by the Instituto de
Astrofisica de Andalucia (IAA) and operated at the Nordic Optical
Telescope under agreement between IAA and the Astronomical Observatory
of the University of Copenhagen.}

\altaffiltext{2}{Institute of Theoretical Astrophysics, University of Oslo, Pb.\ 1029, Blindern, N--0315 Oslo, Norway}
\altaffiltext{3}{Centre for Advanced Study, Drammensvn.~78, N--0271 Oslo, Norway}
\altaffiltext{4}{Astronomical Observatory, University of Copenhagen, Juliane Maries Vej 30, DK--2100 Copenhagen \O, Denmark}
\altaffiltext{5}{NORDITA, Blegdamsvej 17, DK--2100 Copenhagen \O, Denmark}
\altaffiltext{6}{Science Institute, University of Iceland, Dunhaga~3, IS--107 Reykjavik, Iceland}
\altaffiltext{7}{Astronomy Division, Department of Physical Sciences, University of Oulu, FIN-90014 Oulu, Finland}
\altaffiltext{8}{Nordic Optical Telescope, Ap.~474  St.~Cruz de La Palma, E--38700 Canarias, Spain}
\altaffiltext{9}{Isaac Newton Group, Santa Cruz de La Palma, E-38700 Canarias, Spain}
\altaffiltext{10}{Instituto Tecnologie e Studio Radiazioni Extraterrestri, C.N.R., via Gobetti 101, \\ 40129 Bologna, Italy}

\begin{abstract}
We present extensive multi-colour ($UBVR_CI_C$) photometry of the
optical afterglow of GRB~980519. Upon discovery, 8.3~hours after the
burst, the source was decaying as a power law, $(t-t_{\rm
GRB})^\alpha$, with a rapid decay rate $\alpha_1 = - 1.73\pm0.04$.
About 13 hours after the burst a steepening of the light-curve
to $\alpha_2 = -2.22\pm0.04$ was observed.  Within the framework of
current afterglow models, we argue that the rapid initial decline, the
`break' in the light curve, and the spectral properties of the
afterglow are best interpreted as being due to a collimated
ultra-relativistic jet of fixed opening angle expanding into an
inhomogeneous medium.  In this scenario, we find that the circumburst
medium has a density structure that goes as $r^{-2.05\pm0.22}$. This
is characteristic of a preexisting wind expelled from a massive
star. A possible physical scenario is that the progenitor star
collapsed to form a black hole (i.e., a `collapsar'), producing the
observed burst and afterglow. However, the supernova signature
expected in the light curve in such a scenario is not detected. This
either implies that the redshift of GRB~980519 is greater than 1.5 or
that supernovae accompanying GRBs are not standard candles.
\end{abstract}



\section{Introduction}
\label{sec:intro}

The nature of gamma-ray bursts (GRBs) has remained elusive for three
decades. With their cosmological origin firmly established, the
challenge now lies in identifying the GRB progenitors.  Two types of
candidates stand out as the currently most physically viable; the
`collapsing star' and `merging binary compact object' models. These
models can be constrained by the observed temporal evolution and
spectral properties of the GRB afterglow.

The study of GRB afterglow light curves has shown that they are
usually characterized by a power-law decay, $(t-t_{\rm GRB})^\alpha$,
as predicted in the generic ``fireball'' model -- in which the
kinetic energy of ultra-relativisitc particles is converted to photons
from a spatially small region \markcite{piran99}(Piran 1999). A number of afterglows
have been observed to decay at a very rapid rate.  The decay rate,
$\alpha$, depends on the nature of the fireball and also on the
density structure of the ambient medium. There are currently two
generic afterglow scenarios that result in steep light curves; the
jet-model \markcite{rhoads99,sari99}(Rhoads 1999; Sari, Piran, \& Halpern 1999) in which the relativistic outflow is
collimated, and the wind model \markcite{chevalier99}(Chevalier \& Li 1999) which invokes an
inhomogeneous circumburst medium. In a few cases the rate of decay has
been observed to steepen, resulting in a `break' in the light curve,
which is most naturally accounted for by a collimated outflow.

\markcite{halpern99}Halpern {et~al.} (1999) documented the steep power-law decay of GRB~980519
from a collection of own observations of the optical afterglow (OA)
and data reported in circulars. The reported rapid decay, $\alpha =
-2.05\pm0.04$, was interpreted in the context of both a spherical expansion 
into an inhomogeneous medium and a jet-model with lateral expansion in a 
homogeneous medium.  On the basis of the available optical and X-ray data 
\markcite{intZand99,nicastro99}(in't Zand {et~al.} 1999; Nicastro {et~al.} 1999), \markcite{sari99}Sari {et~al.} (1999) advocated a jet-model while 
\markcite{chevalier99}Chevalier \& Li (1999) advocated a wind-model.  \markcite{frail99}Frail {et~al.} (1999) presented radio 
data which favor the wind-model, but could not exclude the jet-model.  
Further observations were reported by \markcite{vrba2000}{Vrba} {et~al.} (2000) and \markcite{smith99}Smith {et~al.} (1998) 
who presented optical and sub-millimeter observations of the OA,
respectively. Finally, \markcite{sokolov98}Sokolov {et~al.} (1998) and \markcite{bloom98b}Bloom {et~al.} (1998b)
independently detected an $R = 26.1\pm0.3$ extended object at the
approximate position of the OA in late July 1998, presumably the host
galaxy.

Here we present $UBVR_CI_C$ photometry of the OA of GRB~980519. We
find that the decay is not described by a single power law as
previously suggested, but steepens about 13 hours after the burst. We
discuss the implications of this result in the context of theoretical
afterglow models.

\section{Observations}
\label{sec:obs}

The OA of GRB~980519 was discovered \markcite{jaunsen98,hjorth98}(Jaunsen {et~al.} 1998; Hjorth {et~al.} 1998) using
the 2.56-m Nordic Optical Telescope (NOT) following a BATSE Rapid
Burst Response on 1998 May 19.51403 UT \markcite{muller98}(Muller 1998), and a
subsequent localisation by the BeppoSAX Wide Field Cameras (WFC) to an
error radius of $3\arcmin$ \markcite{piro98}(Piro 1998).  The first optical images
were inspected and compared to a corresponding Digitized Sky Survey
(DSS)
\footnote{The Digitized Sky Surveys were produced at the Space
Telescope Science Institute under U.S.  Government grant NAG
W-2166. The images of these surveys are based on photographic data
obtained using the Oschin Schmidt Telescope on Palomar Mountain and
the UK Schmidt Telescope.  The plates were processed into the present
compressed digital form with the permission of these institutions.}
image about 8.3~hours after the burst.  The OA was confirmed to be
fading by approximately $0.2$ mag/hour and was observed extensively
during the night of discovery and at regular intervals the
following days, weeks and months. Composite images of the fading
OA are presented in Fig.~\ref{fig:colimg}.

\placefigure{fig:colimg}

An astrometric plate solution computed relative to the USNO--A2.0
catalogue \markcite{monet}(Monet 1998) using the WCS-tools package \markcite{mink}(Mink 1999) gave
$\alpha(J2000)=23^{\rm h}22^{\rm m}21\fs55,
\delta(J2000)=77\arcdeg15\arcmin43\farcs2 $ for the OA.

The images were processed using standard tools in IRAF\footnote{IRAF
(Image Reduction and Analysis Facility) is distributed by the National Optical Astronomy Observatories, which are operated by the Association of Universities for Research in Astronomy, Inc., under contract with the National Science Foundation.}
and the $I_C$-band frames were also de-fringed using a master-fringe image
constructed from all $I_C$-band science images.  On 1998 May 24 we
obtained multi-band imaging of the GRB-field and of the M92-field
centered on the coordinates given in \markcite{christian85.1}Christian {et~al.} (1985) at three
different airmasses for the purpose of independent photometric
calibration. As reference we used a catalogue by P. Stetson
(provided by F. Grundahl) calibrated to the Landolt-system and
measured with PSF photometry.  After matching and pruning the
catalogue covering our sub-field for close neighbors and large
instrumental magnitude errors ($>$ 0.02) we achieved a very good
simultaneous fit of the zero-point, color-term and extinction.  The
RMS and the number of data points used in the fits were 0.023 (36),
0.032 (612), 0.022 (864), 0.025 (862) and 0.026 (553) for the $U$,
$B$, $V$, $R_C$, and $I_C$ bands, respectively. This allowed us to
photometrically calibrate several reference stars in the GRB field for
all five bands to high precision.  These measurements were made using
SExtractor v.2.1.6 \markcite{bertin96}({Bertin} \& {Arnouts} 1996) with fixed apertures plus aperture
corrections extending to 15 arc-seconds.  The M92 photometric
calibration fit and the conversion from instrumental to absolute
magnitudes was done using tools in the IRAF/PHOTCAL package.  In
Table~\ref{tab:fieldphot} we give the results of the field photometry
for all reference objects mentioned in relevant circulars
\markcite{jaunsen98,bloom98a,henden98}(Jaunsen {et~al.} 1998; Bloom {et~al.} 1998a; Henden {et~al.} 1998) using the same nomenclature (two
additional objects introduced by A. Henden are also listed).  Our
absolute photometry of these point sources is in good agreement with
that of \markcite{henden98}Henden {et~al.} (1998) and the robust average deviations are 0.03,
0.03, 0.01, and $0.03$ mag for the $BVR_CI_C$ bands.  A systematic
offset of $0.04$ mag in the $R_C$-band is, however, found.

\placetable{tab:fieldphot}

The OA was found to be consistent with a point source and we therefore
chose to perform relative PSF photometry of the OA and a few
neighboring stars to maximize the signal-to-noise. The PSF photometry
measurements were done using DAOPHOT II \markcite{stetson91}({Stetson} 1991). Absolute
magnitudes were computed for each epoch from the average offset of the
measured stars relative to the aperture measurements described above.
The resulting photometry is listed in Table~\ref{tab:obslog}.

\placetable{tab:obslog}

\section{The lightcurve}
\label{sec:lightcurve}

The magnitudes listed in Table~\ref{tab:obslog} were converted to
$\mu$Jy using the AB offsets and coefficients for the effective
wavelengths of the respective passbands given in \markcite{fukugita95}Fukugita, Shimasaku, \& Ichikawa (1995).
The multi-colour light curves are depicted in
Fig.~\ref{fig:lightcurve}.

\placefigure{fig:lightcurve}

There were no obvious colour changes of the OA throughout the
observing period.  We observed a marked leveling off of the light
curve in the $R_C$-band on 1998 May 30.00 UT. Such a behavior is
usually ascribed to the constant contribution from a host galaxy,
consistent with the detection of a faint, $R = 26.1\pm0.3$, extended
object \markcite{sokolov98,bloom98b}(Sokolov {et~al.} 1998; Bloom {et~al.} 1998b). This host contribution was
subtracted from the combined light curve in the lower panel of
Fig.~\ref{fig:lightcurve}.

A power-law decay of the optical flux, $F_\nu \propto
\nu^{-\beta}t^\alpha$, is expected in fireball models \markcite{sari98}(Sari, Piran, \& Narayan 1998).
Fitting a single power-law, including band offsets relative to the
$R_C$-band was computed by weighted least chi-square minimization
($\chi^2_{25} = 2.4$, where the subscript represents the number of
degrees of freedom in the fit). The errors for each component of the
fit (including the band offset errors) are estimated as the
perturbation needed to give a change in $\chi^2$ of 1 (corresponding
to a 1-sigma error).  The resulting power-law slope is $\alpha_0 =
-2.03 \pm 0.02$, consistent with that found by \markcite{halpern99}Halpern {et~al.} (1999) and
\markcite{vrba2000}{Vrba} {et~al.} (2000).

The combined multi-colour light curve, however, reveal the presence of
a temporal break. A broken power-law fit ($\chi^2_{23} = 0.26$) to the
data gives $\alpha_1 = -1.73\pm0.04$ and $\alpha_2 = -2.22\pm0.04$,
with the break occurring at about $t_{\rm break} \simeq 0.55$ days
after the burst.  The break point was estimated by extrapolating the
two power-laws.  The break in the light curve is highly significant.
Taking the ratio of the two $\chi^2$ values yields $F_{23,25} =
{\chi^2_{25}}/{\chi^2_{23}} = 8.6$.  Assuming the errors to be
Gaussian distributed, $F_{23,25}$ can be approximated by a Fisher
distribution \markcite{lupton93}(Lupton 1993). The broken power-law fit is therefore a
significantly better representation of the light curve than the single
power-law fit at the 99.99995\% confidence level.  The scatter in the
combined multi-band light curve as compared to the broken power-law is
$\lesssim 0.03$ mag. We also experimented with fitting a
four-parameter function to the light-curve, similar to that used in
eg. \markcite{stanek99}{Stanek} {et~al.} (1999), but found the resulting errors much larger than
the two power-law fit.  We estimate the timescale on which the break
occurs to be $\sim0.2$ days (essentially the time difference between
the adjacent end points on the two power laws in a given band).

The X-ray counterpart was observed by the BeppoSAX Wide Field Camera
(WFC) up to about $130$ seconds after the burst trigger.  Follow-up
Narrow Field Instruments (NFI) observations started 10 hours
later. \markcite{nicastro99}Nicastro {et~al.} (1999) attempted to estimate the X-ray decay slope,
$\alpha_X$, based on the WFC and NFI measurements, but noted that the
NFI measurements do not seem to follow a simple power-law decay.  We
interpret their data as the signature of a break in the X-ray decay
slope. The first NFI observation was made prior to the time of our
estimated break, while the remaining NFI observations were made after
the break. Consequently, the pre-break value of $\alpha_X$ can be
estimated as the slope between the WFC and the first NFI measurement,
giving approximately $\alpha_X = -1.6$.
The post-break value of $\alpha_X$ is given approximately by the
minimum estimate of \markcite{nicastro99}Nicastro {et~al.} (1999), $\alpha_X = -2.25\pm0.22$. The
temporal break detected in the optical data is thus independently
confirmed by the X-ray data.

\section{Spectral properties}
\label{sec:specprop}

GRB afterglow is believed to be synchrotron emission by electrons
accelerated at the relativistic shock between the fireball and the
ambient medium, to a power-law energy distribution, $N(\gamma) =
\gamma^{-p}$, where $\gamma$ is the electron Lorentz factor
\markcite{piran99}(Piran 1999).  The resulting spectrum is then characterized by four
power-law sections separated by three breaks at $\nu_{\rm a}$,
$\nu_{\rm m}$ and $\nu_{\rm c}$ corresponding to the synchrotron
self-absorption, peak energy and cooling frequencies, respectively
\markcite{sari98}(Sari {et~al.} 1998).  An optical broad-band spectrum coinciding in time
with the first BeppoSAX NFI observation \markcite{nicastro98}(Nicastro 1998) was
constructed by interpolating (or mildly extrapolating) the flux in
each band to a common date of May 19.96 UT.

Unfortunately, the reddening towards GRB~980519, $b^{II} \sim
15\arcdeg$, is uncertain. The estimate of \markcite{schlegel98}Schlegel, Finkbeiner, \& Davis (1998) for this
region is E(B$-$V) $= 0.267$, but \markcite{arce99}{Arce} \& {Goodman} (1999) argue that these
reddening values are overestimated when $A_V > 0.5$ (which is the case
here). Since a small error in E(B$-$V) translates into a large
uncertainty in the inferred spectral slope, $\beta_O$, we refrain from
estimating it directly from the broad-band spectrum.  Instead, based
on our data, we note that $\beta_O$ can be expressed as a linear
function of the reddening as $ \beta_O \simeq 3.32 $E(B$-$V)$ - 1.60
$.

\placefigure{fig:broadbandspec}

Fig.~\ref{fig:broadbandspec} shows the optical data corrected for
Galactic extinction using E(B$-$V) $= 0.24$ (see discussion in
Section~\ref{sec:model}) and the estimated X-ray spectral index
\markcite{nicastro99}(Nicastro {et~al.} 1999), tied to the first NFI measurement. It is readily
seen from both the uncorrected and extinction corrected optical data
in Fig.~\ref{fig:broadbandspec}, that the $B$ and to some extent $U$
values are inconsistent with a power-law. Assuming the spectral slope
to be a power-law, the flux deficit may be due to extinction in the host
galaxy, depletion by intergalactic Lyman-$\alpha$ absorption, or
both. We therefore fit a power-law spectrum to the $VRI$-bands only.

The difference in the spectral indices of the optical and X-ray
regimes as discussed above and seen in Fig.~\ref{fig:broadbandspec}
indicates the presence of a spectral break at approximately $\nu_c
\simeq 2.3 \times 10^{16}$ Hz.  The steep initial temporal evolution
of GRB~980519 and a spectral break between optical and X-ray suggests
emission by adiabatic electrons ($\nu_{\rm m} < \nu_{\rm c}$).

\section{The model}
\label{sec:model}

When the emitting electrons are adiabatic, it follows that the initial
temporal evolution of the light-curve (whether the outflow is
spherical or collimated) is given by
$
F_\nu\propto\nu^{-\beta}t^{-3\beta/2-\delta/(8-2\delta)}
$,
with the ambient density profile of the form $n(r)\propto r^{-\delta}$
($\delta < 3$) \markcite{meszaros98,pana98}(M\'esz\'aros, Rees, \&  Wijers 1998; Panaitescu, M\'esz\'aros, \& Rees 1998). Here, $\delta=2$, represents
a stellar wind density distribution, while $\delta=0$ represents a
homogeneous density distribution.  The temporal decay slope is thus
steeper by $\delta/(8-2\delta)$ in the inhomogeneous case.

If the ejecta is collimated, and the observer is within the collimated
jet, the light curve is given by the above expression while there
still is considerable relativistic beaming, i.e.\ while $\Gamma >
1/\theta$, where $\Gamma$ is the bulk Lorentz-factor and $\theta$ is
the half angle of the collimated outflow.  When the expansion slows
down and $\Gamma$ drops below $1/\theta$, the observer starts to see
the edge of the jet and the light curve behavior switches to
$
F_\nu\propto \nu^{-\beta}t^{-3\beta/2-(6-\delta)/(8-2\delta)}.
$
The light curve thus steepens by $\Delta \alpha = (3-\delta)/(4-\delta$). An
inhomogeneous environment therefore leads to a steeper light curve
but a less pronounced break than in the homogeneous case. 
Specifically, a collimated outflow expanding into a circumstellar wind 
environment with $\delta=2$, would steepen by $1/2$, whereas the steepening 
in a homogeneous environment would be $3/4$. 

In the case of GRB~980519 the existence of a temporal break is most
naturally explained if the outflow was collimated. The steepening in
the light curve is measured to be $\Delta \alpha = \alpha_1 - \alpha_2
= 0.49 \pm0.06$ which implies an environmental density gradient of
$\delta=2.05 \pm 0.22$, in excellent agreement with the distribution
expected from a circumstellar wind.  The expected optical spectral
index can be inferred from the light curve as $\beta = -2/3
(\alpha_1+\delta/(8-2\delta)) = 0.80 \pm 0.08$.  The relation between
reddening due to Galactic extinction and $\beta_O$ given in
Section~\ref{sec:specprop} can then be used to infer E(B$-$V) $= 0.24$
from the expected optical spectral index.  This is the value used for
the Galactic extinction correction of the optical data in
Fig.~\ref{fig:broadbandspec} and is in good agreement with the value
estimated from \markcite{schlegel98}Schlegel {et~al.} (1998).  The power-law index of the electron
energy distribution is $p = 1+ 2\beta = 2.61\pm0.16$ (for $\nu_{\rm m}
< \nu < \nu_{\rm c}$).  The X-ray spectral index then is $\beta_X
=p/2= 1.30\pm0.08$ (for $\nu > \nu_{\rm c}$), consistent with the
estimate of \markcite{nicastro99}Nicastro {et~al.} (1999), $\beta_X = 1.8^{+0.6}_{-0.5}$, and that
of \markcite{owens98}{Owens} {et~al.} (1998), $\beta_X = 1.52^{+0.7}_{-0.57}$.

\section{Discussion}
\label{sec:conclusions}

In the 'collapsar' scenario the gamma-ray burst and its afterglow are
believed to be produced by a collapsing massive star (e.g., a
Wolf--Rayet star) leading to a collimated ultra-relativistic
outflow. The jet would then interact with an inhomogeneous density
distribution created by a preexisting stellar wind expelled by the
progenitor star \markcite{fryer99,macfadyen99}(Fryer, Woosley, \& Hartmann 1999; MacFadyen \& Woosley 1999).

The identification of a spectral break between optical and X-rays,
suggesting a fast cooling regime ($\nu > \nu_{\rm c}$) at X-ray
frequencies, implies that the X-ray decay slope is shallower by $1/4$
as compared to the optical (slow cooling) \markcite{chevalier99}(Chevalier \& Li 1999).  This
is consistent with the temporal X-ray decay discussed in
Section~\ref{sec:lightcurve}.  The radio data presented by
\markcite{frail99}Frail {et~al.} (1999) also appears to be consistent with the presented
collapsar scenario.  The collapsar model can thus account for the
X-ray, optical and radio observations of GRB~980519.

A potential problem for this interpretation concerns the sharpness of
the observed break.  The break in the light curve of GRB~980519 is
rather sharply defined in time and occurred on a timescale of about
0.2 days. This is much shorter than expected from theoretical models.
\markcite{moderski2000}{Moderski}, {Sikora}, \&  {Bulik} (2000) showed that the transition of the light curve
from one asymptote to the other occurs within a factor of two in time
for a jet of constant opening angle in a constant density
environment. If the opening angle of the collimated outflow is allowed
to evolve, the break becomes even smoother, i.e., extends over a
longer period of time.  Based on numerical simulations,
\markcite{kumar2000}Kumar \& Panaitescu (2000) argue that the transition in such cases takes place
over a factor of at least 100 in time in a wind density
environment. Much shorter break time scales, however, are expected for
observers located on the jet axis.

In the collapsar scenario a re-brightening of the source $\sim 15
(1+z)$ days after the burst is expected due to the onset of a
supernova type Ibc explosion as energy is deposited by the jet
\markcite{macfadyen99}(MacFadyen \& Woosley 1999).  \markcite{bloom99}Bloom {et~al.} (1999) noted that the rapid decay of the
OA of GRB~980519 and faintness of the host makes it a good candidate
for a SN detection. The expected late evolution based on the SN~1998bw
light curves associated with GRB 980425 \markcite{galama98}(Galama {et~al.} 1998) are therefore
given in Fig.~\ref{fig:lightcurve} at various redshifts as in
\markcite{bloom99}Bloom {et~al.} (1999). We do not detect a re-brightening comparable to the
SN~1998bw light curves.  However, our last detection of the
afterglow+host on 1998 May 30.00 UT, the subsequent upper limits and
host detections by \markcite{sokolov98,bloom98b}Sokolov {et~al.} (1998); Bloom {et~al.} (1998b) set a lower limit to the
redshift of GRB~980519 of $z \gtrsim 1.5$ assuming the expected
re-brightening is similar to SN~1998bw. This constraint on the
redshift is consistent with the faintness of the host galaxy.

Alternatively, should the redshift of GRB~980519 turn out to be lower
than $1.5$, the absence of a SN signature in the data (see
Fig.~\ref{fig:lightcurve}) can be taken as evidence that collapsars do
not necessarily produce SNe of comparable brightness (`standard
candle') to SN~1998bw \markcite{hjorth2000}(see Hjorth {et~al.} 2000, suggesting this is the case in
GRB~990712).

Finally, we note that \markcite{esin2000}Esin \& Blandford (2000) recently presented an
alternative explanation for the excess red flux observed $\sim$ 20 --
30 d after the bursts of GRB 970228 \markcite{fruchter99}({Fruchter} {et~al.} 1999) and GRB~980326
\markcite{bloom99}(Bloom {et~al.} 1999), hitherto taken as evidence for the SNe interpretation
and indirectly a 'collapsar' origin. Future observations should allow
a discrimination between the two models.


\acknowledgments

\noindent ACKNOWLEDGEMENTS.  
We thank L. Nicastro for contributing updated and refined X-ray
values.  AOJ thanks S. V. Haugan and O. Skj\ae raasen for helpful
discussions.  We thank the NOT Director for continued support to our
GRB programme. This research was supported by the Danish Natural
Science Research Council (SNF), the Icelandic Council of Science and
the University of Iceland Research Fund.

\clearpage

\begin{deluxetable}{lrrrrrrr}
\tabletypesize{\scriptsize}
\tablecaption{$UBVR_CI_C$ field photometry \label{tab:fieldphot}.}
\tablewidth{0pt}
\tablehead{
\colhead{ID} & \colhead{RA$_{\rm J2000}$} & \colhead{DEC$_{\rm J2000}$} & \colhead{$U$} & \colhead{$B$} & \colhead{$V$} & \colhead{$R_C$} & \colhead{$I_C$}
}
\startdata
A & 23:22:17.741 & +77:15:51.199 &    20.42 $\pm$     0.03&   19.62 $\pm$     0.01&   18.54 $\pm$     0.01&   17.88 $\pm$     0.01&   17.34 $\pm$     0.01 \nl
B & 23:22:27.290 & +77:15:43.499 &    22.78 $\pm$     0.11&   21.59 $\pm$     0.05&   20.21 $\pm$     0.02&   19.27 $\pm$     0.01&   18.46 $\pm$     0.01 \nl
C & 23:22:28.500 & +77:15:25.402 &    19.49 $\pm$     0.03&   19.36 $\pm$     0.01&   18.52 $\pm$     0.01&   17.99 $\pm$     0.01&   17.52 $\pm$     0.01 \nl
D & 23:22:47.141 & +77:16:23.498 &    16.75 $\pm$     0.03&   16.42 $\pm$     0.00&   15.60 $\pm$     0.00&   15.10 $\pm$     0.00&   14.68 $\pm$     0.00 \nl
E & 23:22:20.201 & +77:15:24.001 &  \nodata& \nodata&   21.15 $\pm$     0.04&   19.97 $\pm$     0.02&   18.63 $\pm$     0.02 \nl
F & 23:22:31.399 & +77:15:43.301 &  \nodata&   22.63 $\pm$     0.12&   20.98 $\pm$     0.04&   19.81 $\pm$     0.02&   18.26 $\pm$     0.01 \nl
G & 23:22:33.199 & +77:14:58.999 &    22.74 $\pm$     0.10&   21.42 $\pm$     0.04&   19.88 $\pm$     0.02&   18.86 $\pm$     0.01&   17.93 $\pm$     0.01 \nl
H & 23:22:20.700 & +77:14:55.000 &    22.86 $\pm$     0.12&   22.05 $\pm$     0.07&   20.52 $\pm$     0.03&   19.39 $\pm$     0.01&   17.97 $\pm$     0.01 \nl
I & 23:22:09.199 & +77:15:51.001 &    20.60 $\pm$     0.04&   20.73 $\pm$     0.03&   20.02 $\pm$     0.02&   19.61 $\pm$     0.01&   19.18 $\pm$     0.02 \nl
J & 23:22:07.250 & +77:16:00.599 &  \nodata&   22.43 $\pm$     0.10&   20.93 $\pm$     0.03&   19.96 $\pm$     0.02&   19.14 $\pm$     0.02 \nl
M & 23:22:46.610 & +77:16:43.399 &    20.17 $\pm$     0.03&   19.10 $\pm$     0.01&   17.93 $\pm$     0.01&   17.21 $\pm$     0.00&   16.60 $\pm$     0.00 \nl
N & 23:22:39.610 & +77:12:46.901 &  \nodata&   16.74 $\pm$     0.01&   15.94 $\pm$     0.01&   15.45 $\pm$     0.01&   15.03 $\pm$     0.01 \nl
\enddata
\tablecomments{World coordinates computed relative to USNO A--2.0 \markcite{monet}(Monet 1998) using WCS-tools \markcite{mink}(Mink 1999)}
\end{deluxetable}

\clearpage

\begin{deluxetable}{lrrrr}
\tabletypesize{\scriptsize}
\tablecaption{Observing log and magnitudes\label{tab:obslog}.}
\tablewidth{0pt}
\tablehead{
\colhead{Date} & \colhead{Exp. time} & \colhead{Band} & \colhead{Seeing} & \colhead{Johnson-Cousins}\\
\colhead{1998 May UT} & \colhead{(sec)} & \colhead{} & \colhead{FWHM (\arcsec)} & \colhead{Magnitude}
}
\startdata
19.861 &    60 & I &  1.20 &    18.42 $\pm$     0.09 \nl
19.864 &   180 & I &  1.33 &    18.47 $\pm$     0.05 \nl
19.877 &   300 & I &  1.27 &    18.52 $\pm$     0.02 \nl
19.933 &   500 & I &  1.79 &    18.83 $\pm$     0.04 \nl
20.002 &   500 & I &  1.51 &    19.09 $\pm$     0.02 \nl
20.047 &   500 & I &  1.49 &    19.26 $\pm$     0.02 \nl
20.136 &   500 & I &  1.02 &    19.60 $\pm$     0.02 \nl
20.210 &   500 & I &  0.92 &    19.89 $\pm$     0.02 \nl
20.976 &   500 & I &  1.39 &    21.71 $\pm$     0.20 \nl
21.164 &   600 & I &  1.21 &    21.88 $\pm$     0.11 \nl
21.889 &  1200 & I &  1.01 &    22.80 $\pm$     0.23 \nl
22.192 &  1200 & I &  1.02 &    23.28 $\pm$     0.18 \nl
19.887 &   300 & V &  1.97 &    19.64 $\pm$     0.02 \nl
19.926 &   300 & V &  2.40 &    20.01 $\pm$     0.04 \nl
19.940 &   500 & V &  2.35 &    19.99 $\pm$     0.03 \nl
19.993 &   500 & V &  1.99 &    20.22 $\pm$     0.02 \nl
20.040 &   500 & V &  1.62 &    20.36 $\pm$     0.02 \nl
20.129 &   500 & V &  1.32 &    20.67 $\pm$     0.02 \nl
20.203 &   500 & V &  1.11 &    20.98 $\pm$     0.02 \nl
20.230 &   450 & V &  1.50 &    20.90 $\pm$     0.10 \nl
20.969 &   500 & V &  1.53 &    22.68 $\pm$     0.12 \nl
21.156 &   500 & V &  1.03 &    23.05 $\pm$     0.11 \nl
20.054 &   500 & R &  1.52 &    19.84 $\pm$     0.01 \nl
20.151 &   500 & R &  1.03 &    20.27 $\pm$     0.01 \nl
20.223 &   300 & R &  1.09 &    20.48 $\pm$     0.03 \nl
23.190 &  5100 & R &  0.84 &    24.31 $\pm$     0.12 \nl
30.003 &  9600 & R &  0.62 &    25.74 $\pm$     0.25 \nl
20.061 &   500 & B &  1.71 &    21.12 $\pm$     0.02 \nl
20.143 &   500 & B &  1.38 &    21.50 $\pm$     0.03 \nl
20.217 &   500 & B &  1.26 &    21.83 $\pm$     0.05 \nl
20.069 &   600 & U &  1.78 &    20.75 $\pm$     0.07 \nl
\enddata
\end{deluxetable}

\clearpage

\figcaption[fig1.ps]{
True-colour images centered on the fading optical counterpart of
GRB~980519 at epoch 1998 May 20.05, 20.136 and 20.21 UT. North is up
and East is to the left. The field measures $1\arcmin \times
1\arcmin$.
\label{fig:colimg}
}

\figcaption[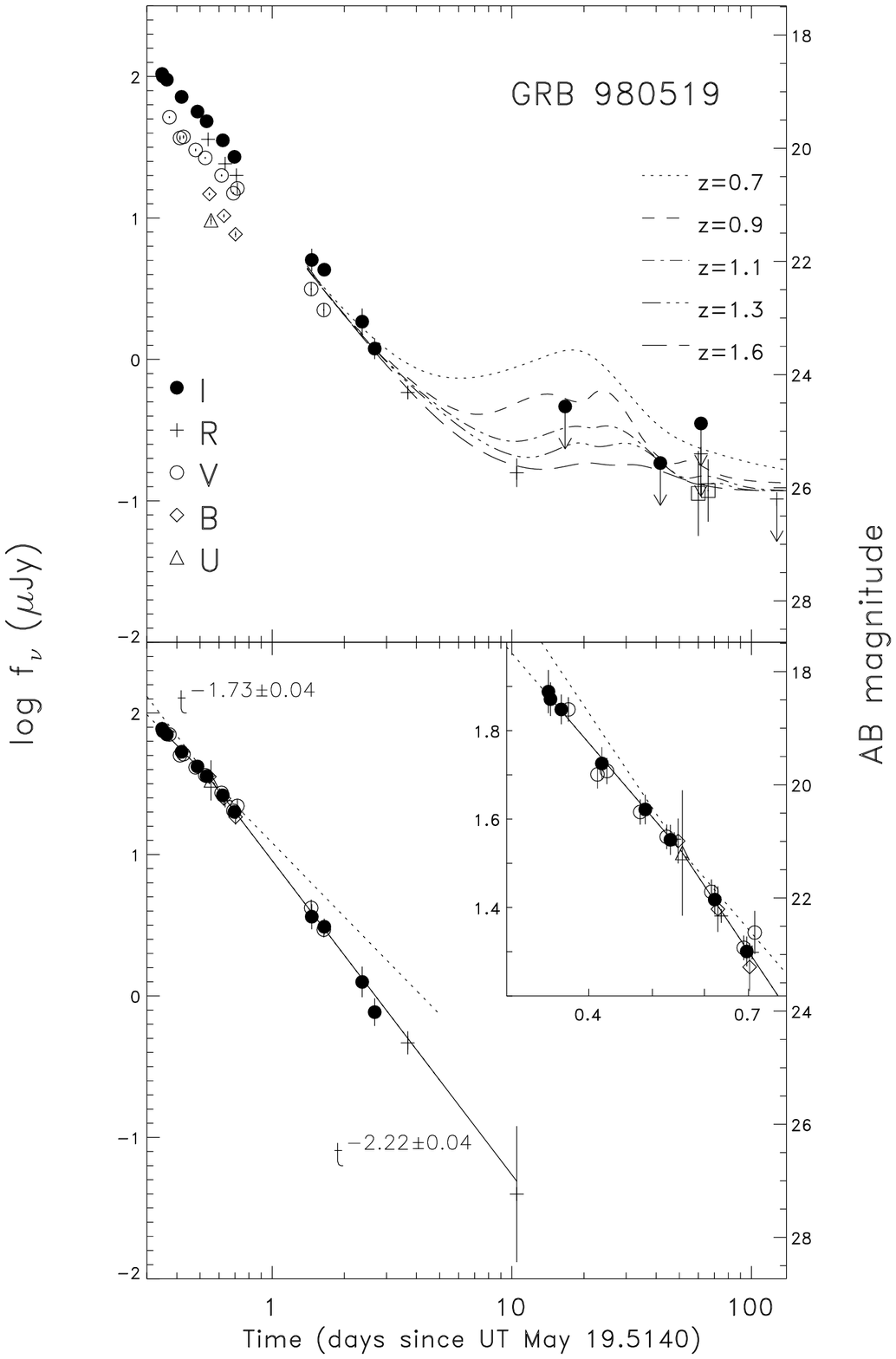]{
The upper panel shows the multi-colour light curves of
GRB~980519 with the expected re-brightening due to an SN explosion
based on the SN~1998bw light curves \markcite{galama98}(Galama {et~al.} 1998).  The host
detections by \markcite{sokolov98}Sokolov {et~al.} (1998) and \markcite{bloom98b}Bloom {et~al.} (1998b) are marked by
boxes.  The lower panel shows the combined data points after the
offsets relative to the $R_C$-band data have been applied and the host
contribution subtracted. The interpolated (solid lines) and
extrapolated (dotted lines) power-laws are also plotted.  The inset
shows the data from the first night only.
\label{fig:lightcurve}
}

\figcaption[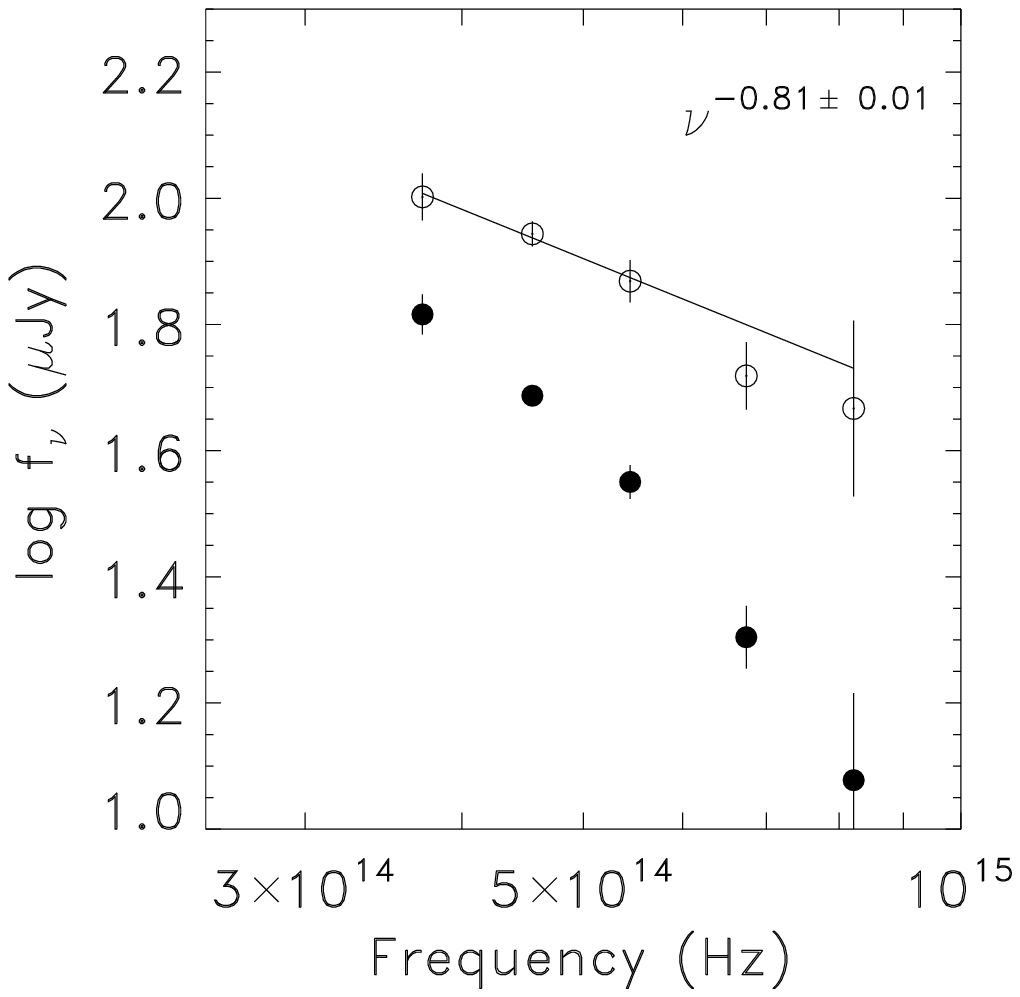,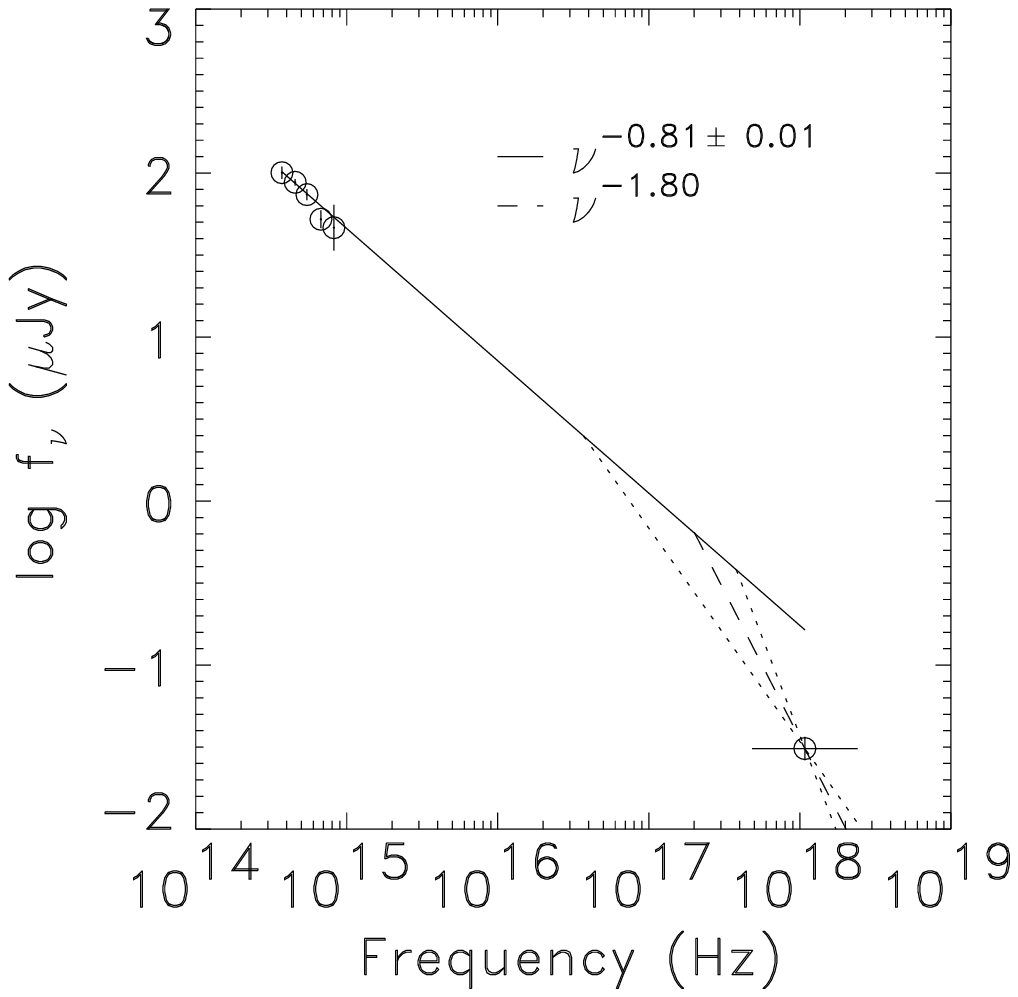]{
In the left panel we show the broad-band spectrum before (filled) and
after (open) correction for Galactic extinction. The right panel shows
the corrected and extrapolated optical spectral slope and the X-ray
measurement by \markcite{nicastro99}Nicastro {et~al.} (1999) with the estimated X-ray slope
corresponding to $\beta_X = 1.8^{+0.6}_{-0.5}$ (dashed) and its
1-sigma uncertainty values tied by the X-ray measurement (dotted).
\label{fig:broadbandspec}
}

\clearpage

\epsscale{1}
\plotone{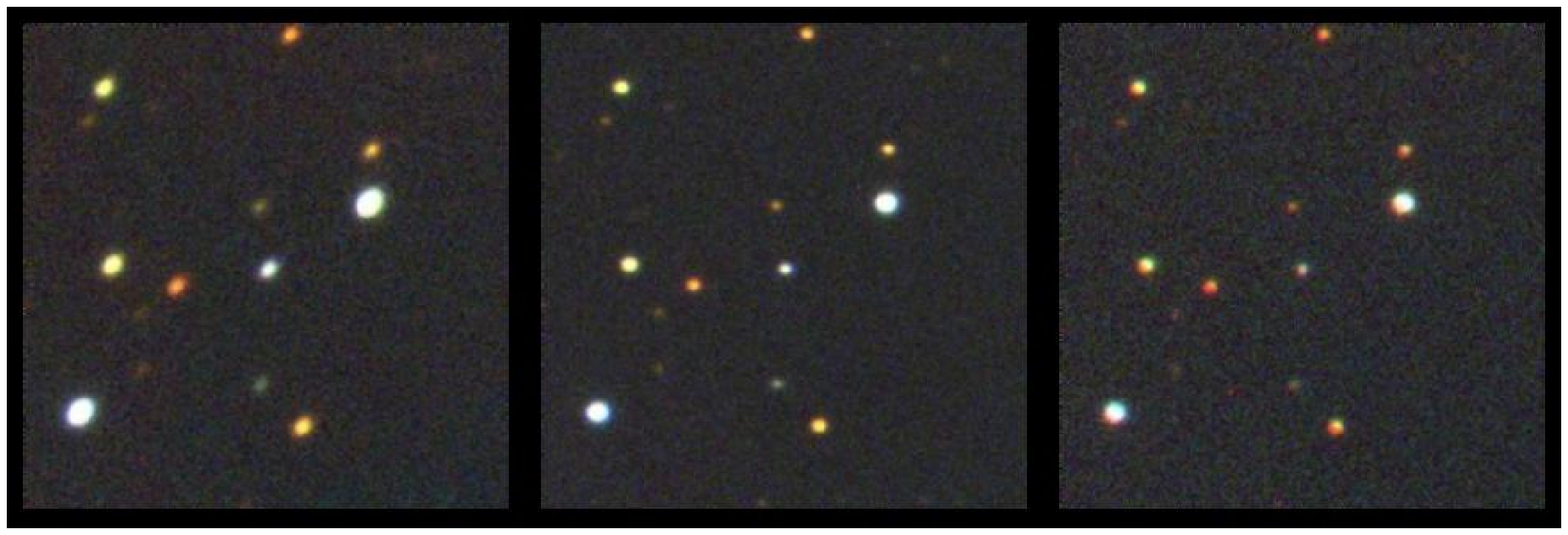}

\clearpage

\epsscale{0.7}
\plotone{fig2.eps}

\clearpage

\epsscale{1}
\plottwo{fig3a.eps}{fig3b.eps}

\end{document}